\documentclass[aps,showpacs,preprintnumbers,twocolumn,amsmath,amssymb,prl,floatfix,superscriptaddress]
{revtex4}

\usepackage{graphicx}
\usepackage[lofdepth,lotdepth]{subfig}
\usepackage{comment}
\usepackage{color}

\usepackage{ifpdf}

\ifpdf
\usepackage{epstopdf}
\usepackage[pdftex]{hyperref}
\else
\usepackage[hypertex,ps2pdf,dvips,colorlinks,urlcolor=blue,citecolor=blue,linkcolor=blue]{hyperref}
\fi
\makeatletter

\newcommand{\vev}[1]{\left\langle #1 \right\rangle}

\def\be{\begin{equation}}
\def\ee{\end{equation}}

\newcommand\mc            {\mathcal}
\newcommand\p             {\partial}

\newcommand\erf           \eqref

\definecolor{Blue}{rgb}{0.00, 0.00, 1.00}
\definecolor{Red}{rgb}{1.00, 0.00, 0.00}

\makeatother

\begin{document}

\title{Correlations after Quantum Quenches in the \textit{XXZ} Spin Chain:\\ Failure
of the Generalized Gibbs Ensemble}

\author{B. Pozsgay}
\affiliation{MTA-BME ``Momentum'' Statistical Field Theory Research Group\\
Department of Theoretical Physics, Budapest University of Technology and Economics, 1111 Budapest, Budafoki \'ut 8, Hungary}
\author{M. Mesty\'an}
\affiliation{MTA-BME ``Momentum'' Statistical Field Theory Research Group\\
Department of Theoretical Physics, Budapest University of Technology and Economics, 1111 Budapest, Budafoki \'ut 8, Hungary}
\author{M. A. Werner}
\affiliation{MTA-BME ``Momentum'' Exotic Quantum Phases Research Group\\
Department of Theoretical Physics, Budapest University of Technology and Economics, 1111 Budapest, Budafoki \'ut 8, Hungary}
\author{M. Kormos}
\affiliation{MTA-BME ``Momentum'' Statistical Field Theory Research Group\\
Department of Theoretical Physics, Budapest University of Technology and Economics, 1111 Budapest, Budafoki \'ut 8, Hungary}
\author{G. Zar\'and}
\affiliation{MTA-BME ``Momentum'' Exotic Quantum Phases Research Group\\
Department of Theoretical Physics, Budapest University of Technology and Economics, 1111 Budapest, Budafoki \'ut 8, Hungary}
\author{G. Tak\'acs}
\thanks{Corresponding author, takacsg@eik.bme.hu}
\affiliation{MTA-BME ``Momentum'' Statistical Field Theory Research Group\\
Department of Theoretical Physics, Budapest University of Technology and Economics, 1111 Budapest, Budafoki \'ut 8, Hungary}

\begin{abstract}

We study the nonequilibrium time evolution of the spin-1/2 anisotropic Heisenberg (\textit{XXZ}) spin chain, with a choice of  dimer product and N\'eel states  as initial states.
We investigate numerically various short-ranged spin correlators in the long-time limit  and find that they deviate significantly from predictions based on the generalized Gibbs ensemble (GGE) hypotheses. By 
computing the asymptotic spin correlators within   the recently proposed quench-action formalism [Phys. Rev. Lett. 110, 257203 (2013)], however, we find excellent agreement with the numerical data. 
We, therefore, conclude  that the GGE cannot give a complete description even of local  observables,
while the quench-action formalism correctly captures the steady state in this case.

\end{abstract}


\maketitle

\paragraph{Introduction.---}
Recent, spectacular advances in the field of ultracold atoms enabled experimentalists to investigate the coherent time evolution of almost perfectly isolated quantum many-body systems  \cite{2006Natur.440..900K,2012Natur.481..484C,2013Natur.502...76F,trotzky}.  These new developments triggered tremendous theoretical interest \cite{2011RvMP...83..863P,kollath-lauchli-altman,2008Natur.452..854R,barthel-schollwock,cramer-eisertNJP,banuls,santosrigolPRE81,barmettler,2011PhRvL.106v7203C,2012JSMTE..07..016C,rigolPRL103,sedlmayr,marcuzziPRL111,andreiPRA87} in a long-standing problem of fundamental physical importance: do isolated quantum systems reach an equilibrium in some sense,
and if the answer is positive, what is the nature of the steady state reached?

In the absence of external driving forces, generic systems
are expected to reach a steady state locally indistinguishable from
thermal equilibrium \cite{2008Natur.452..854R,2011RvMP...83..863P}.
However, integrable systems behave differently because conservation of the
expectation values of extra local charges prevents relaxation to a thermal state. It was suggested in Ref. \cite{2007PhRvL..98e0405R}  that in the integrable case 
the long-time asymptotic stationary state is described by a statistical ensemble involving all the local conserved charges $\{\hat Q_i\},$ the generalized Gibbs ensemble (GGE), defined by the density matrix
\be
\hat \rho_\text{GGE}=\frac{1}{Z}e^{-\sum_{i}\beta_{i}\hat Q_{i}},\qquad Z=\mbox{Tr }e^{-\sum_{i}\beta_{i}\hat Q_{i}}\,,
\label{eq:GGE}
\ee
where the ``chemical potentials'' $\{\beta_i\}$ are determined by the expectation values
of the charges in the ensemble and the initial quantum state. 

The GGE idea has by now become widely accepted in the field and has been verified in several specific cases.
Until recently, however, most investigations concerning GGE were carried out in theories
equivalent to free fermions \cite{cazalillaPRL97,2009PhRvL.102l7204R,2012JSMTE..07..022C,cazalilla-iucci-chung,gurarie,2013PhRvB..87x5107F,2014JSMTE..03..016F,spyrostrap1, kormos2014analytic,spyrosGGE}
or by numerical studies of  relatively small systems \cite{PhysRevLett.106.140405,2013arXiv1312.4657W}.
It is only recently that it has become possible to examine genuinely interacting  integrable systems
such as the one-dimensional Bose gas \cite{caux-konik,2013PhRvB..88t5131K,2014PhRvA..89c3601D}, the \textit{XXZ} Heisenberg spin chain \cite{2013JSMTE..07..003P,2013JSMTE..07..012F,2014PhRvB..89l5101F} or field theories \cite{fm-10,mussardoPRL111,spyros-gabor-giuseppe}. 

However, until now there have only been a few precision numerical tests for the predictions of the GGE
against real-time dynamics  \cite{2014PhRvB..89l5101F}. In our Letter, 
we perform real-time numerical simulations on a genuinely interacting quantum system, the anisotropic Heisenberg model, and compare the relaxation of various local spin-spin correlation functions to
the predictions of two competing theories: the overlap-incorporating
thermodynamic Bethe ansatz (OTBA) approach, which implements the quench-action method
\cite{2013PhRvL.110y7203C}, and the widely accepted GGE.

In agreement with some recent observations~\cite{JStalk,2014arXiv1405.0172W} we find that
 these two approaches  yield markedly different predictions.  We arrive at a surprising conclusion:  while the numerical results agree spectacularly with the OTBA,  they differ significantly from the exact predictions of the GGE in a number of cases (see Fig.\ \ref{fig2}).   These 
results lead to the inevitable conclusion that the GGE approach fails as a generic description of steady states in genuinely interacting integrable quantum systems. 


\begin{figure*}[t]
{\includegraphics[width=0.32\textwidth]{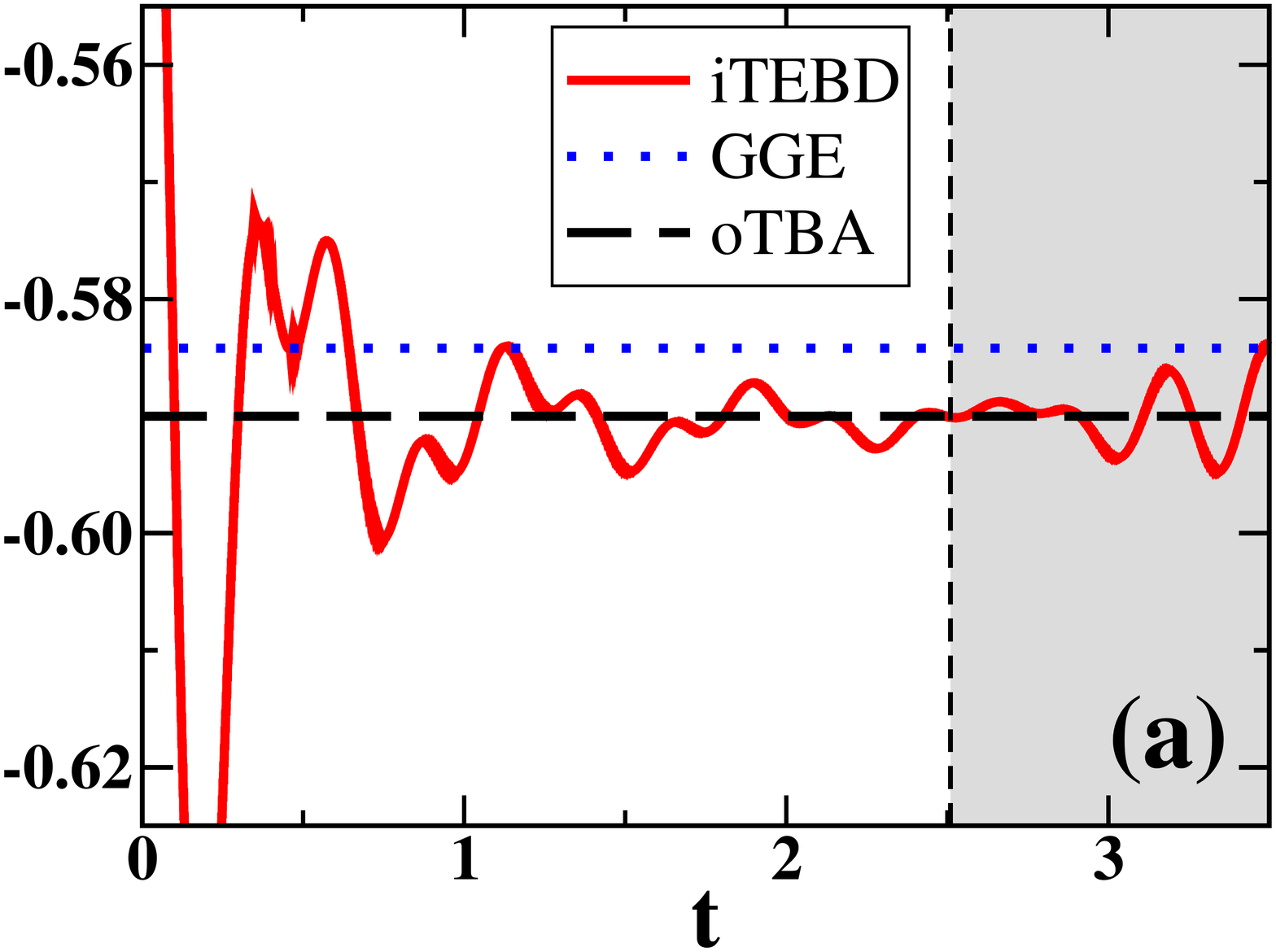}\label{figa}}
{\includegraphics[width=0.32\textwidth]{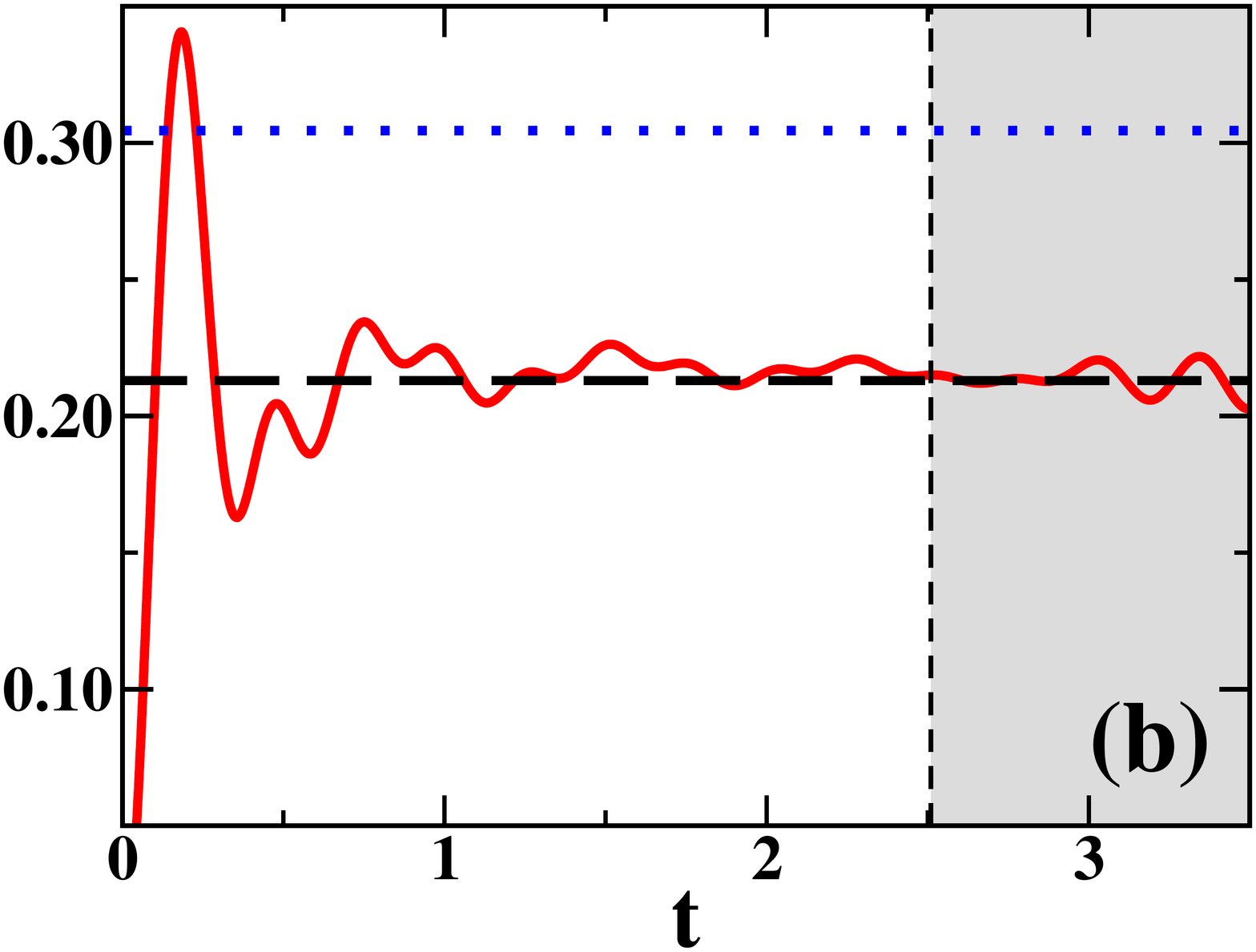}\label{figb}}
{\includegraphics[width=0.32\textwidth]{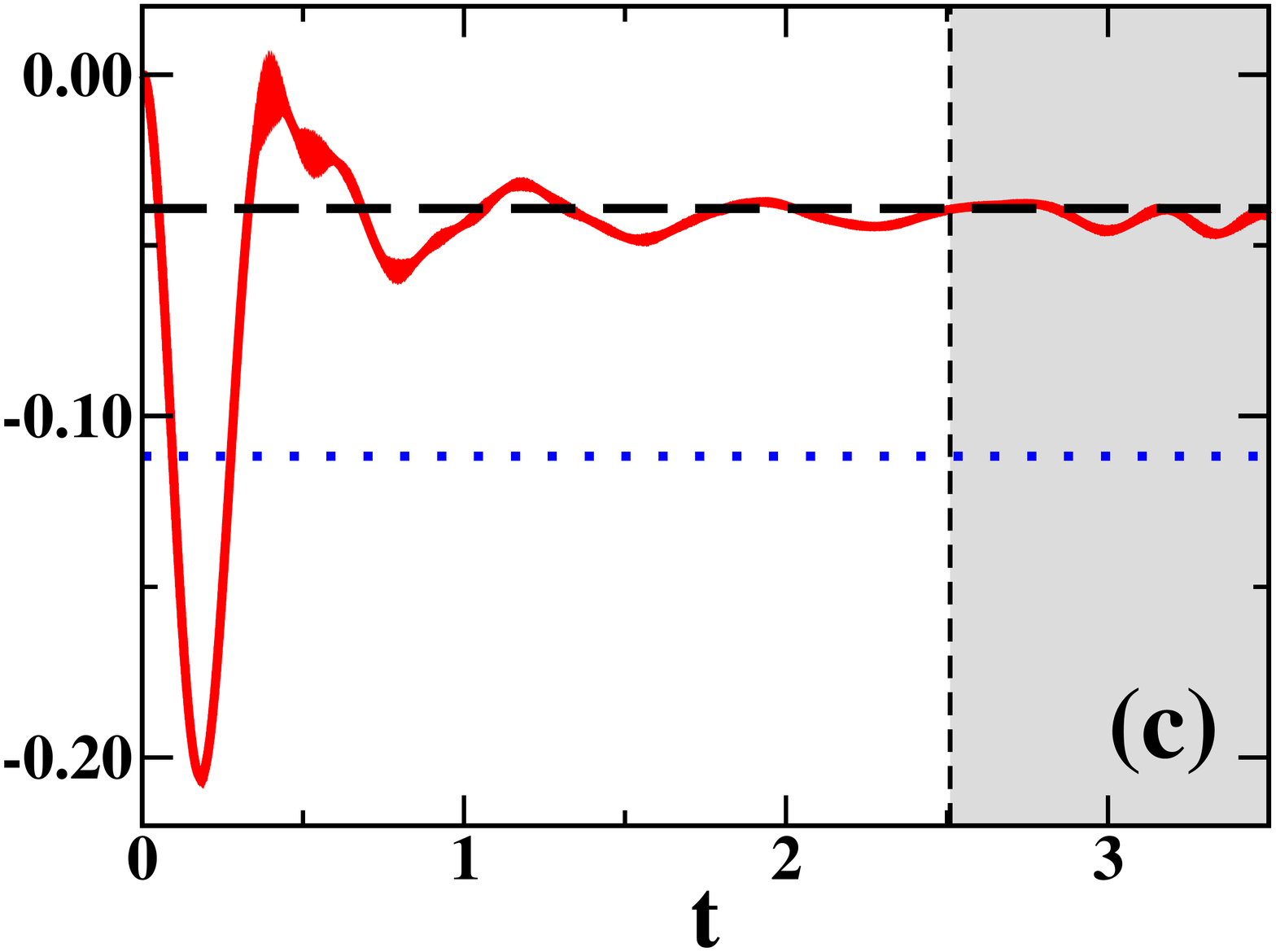}\label{figc}}
\caption{Numerical simulation of the time evolution of correlation functions (a) $\langle\sigma_1^z\sigma^z_2\rangle$, (b) $\langle\sigma_1^z\sigma^z_3\rangle$, (c) $\langle\sigma_1^z\sigma^z_4\rangle$ starting from the dimer initial state \erf{dimer} for anisotropy $\Delta=4$ as obtained by ITEBD (red lines). In the shaded region the results are not reliable. The horizontal lines show the GGE prediction \cite{2014PhRvB..89l5101F} (blue dotted lines) and the prediction of the overlap TBA of the quench-action approach (black dashed lines). } 
\label{fig1}
\end{figure*}


\paragraph{Quantum quench in the XXZ chain.---}

The \textit{XXZ} Heisenberg  chain is a chain 
of $s=1/2$ spins interacting via the Hamiltonian
\be
H=-\sum_{i=1}^L\left[\sigma_{i}^{x}\sigma_{i+1}^{x}+\sigma_{i}^{y}\sigma_{i+1}^{y}+
\Delta (\sigma_{i}^{z}\sigma_{i+1}^{z}-1) 
\label{H}
\right],
\ee
where $\sigma_{i}^{x,y,z}$ are the Pauli matrices at site $i.$ This model describes magnetism in real compounds \cite{assa1994interacting} and plays a fundamental role in the theory of strongly correlated
condensed-matter systems \cite{2004cond.mat.12421E}. Here
we focus on the Ising regime $\Delta>1,$ which corresponds to a gapped antiferromagnetic phase in equilibrium.

We implement the nonequilibrium process  via  the paradigmatic setting of quantum quench \cite{calabrese2005evolution,2006PhRvL..96m6801C},  whereby the time evolution of the system starts from the ground state of some Hamiltonian but then at time $t=0$ some parameter of the system is abruptly
changed. Quantum quenches of this kind can be implemented in a controlled fashion by realizing the \textit{XXZ} chain in systems of cold atoms in optical lattices \cite{PhysRevLett.107.210405,PhysRevA.79.053627,simon2011quantum,PhysRevLett.90.100401,PhysRevLett.91.090402}.

As initial states, we consider the translationally invariant projection of the N\'eel state
\be
|\Psi_0^N\rangle = \frac{1+\hat T}{\sqrt{2}} |\!\uparrow\downarrow \uparrow\ldots\rangle =
\frac{1}{\sqrt{2}}\big(|\!\uparrow\downarrow \uparrow\ldots\rangle+|\!\downarrow\uparrow\downarrow \ldots\rangle\big),
\label{neel}
\ee
which is a ground state in the $\Delta\to\infty$ limit, and the similarly symmetrized dimer product state
\be
|\Psi_0^D\rangle = \frac{1+\hat T}{\sqrt{2}}\, \left|\!\frac{(\uparrow\downarrow-\downarrow\uparrow)}{\sqrt{2}} \frac{(\uparrow\downarrow-\downarrow\uparrow)}{\sqrt{2}} \ldots\right\rangle,
\label{dimer}
\ee
which is one of the ground states of the Majumdar-Ghosh Hamiltonian \cite{MG}. Here, $\hat T$ is the one site translation operator on the lattice. It is expected that ground states of local Hamiltonians always relax to a steady state.


\begin{figure*}
{\includegraphics[width=0.45\textwidth]{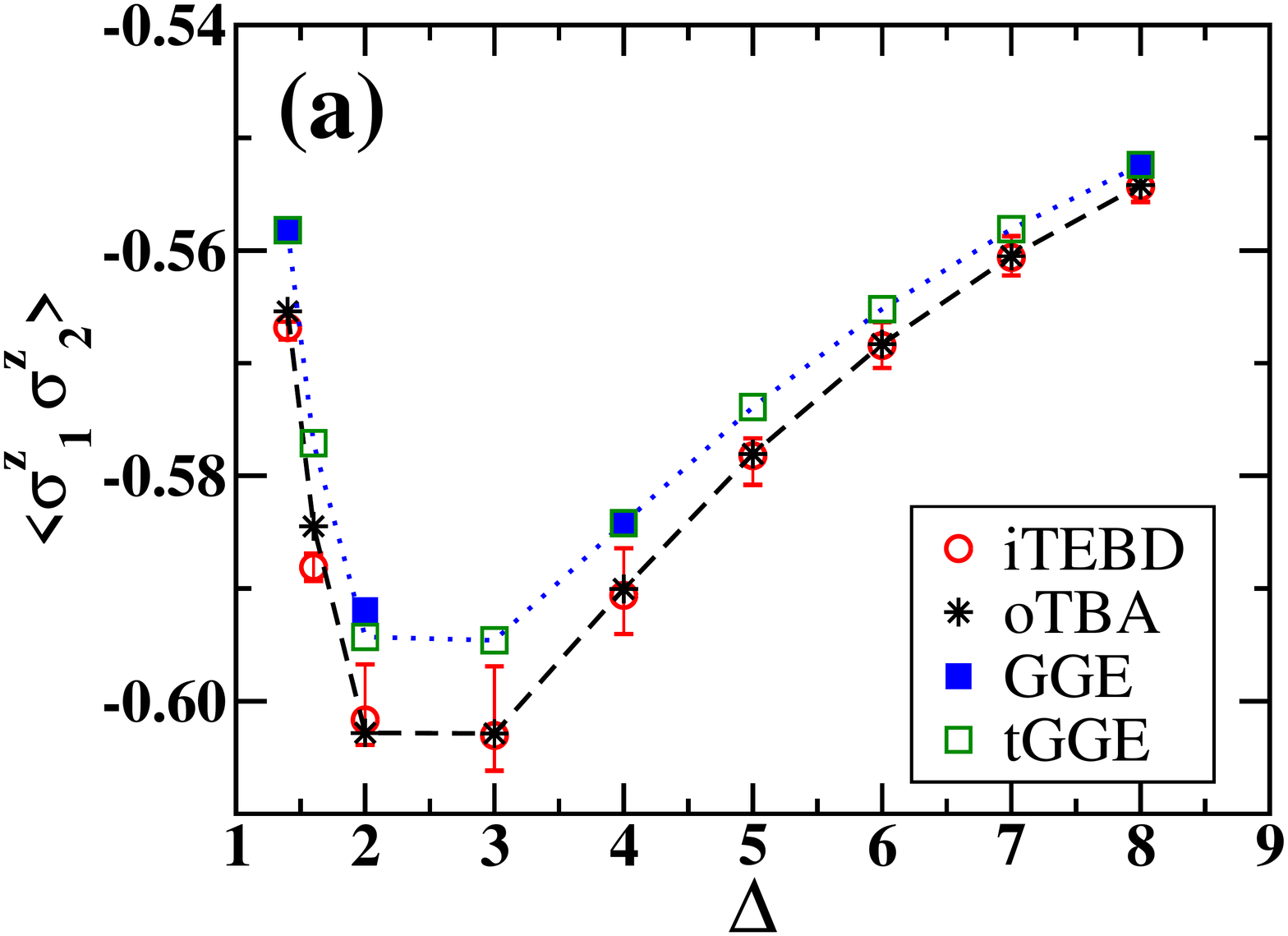}}
\hspace{0.04\textwidth}
{\includegraphics[width=0.45\textwidth]{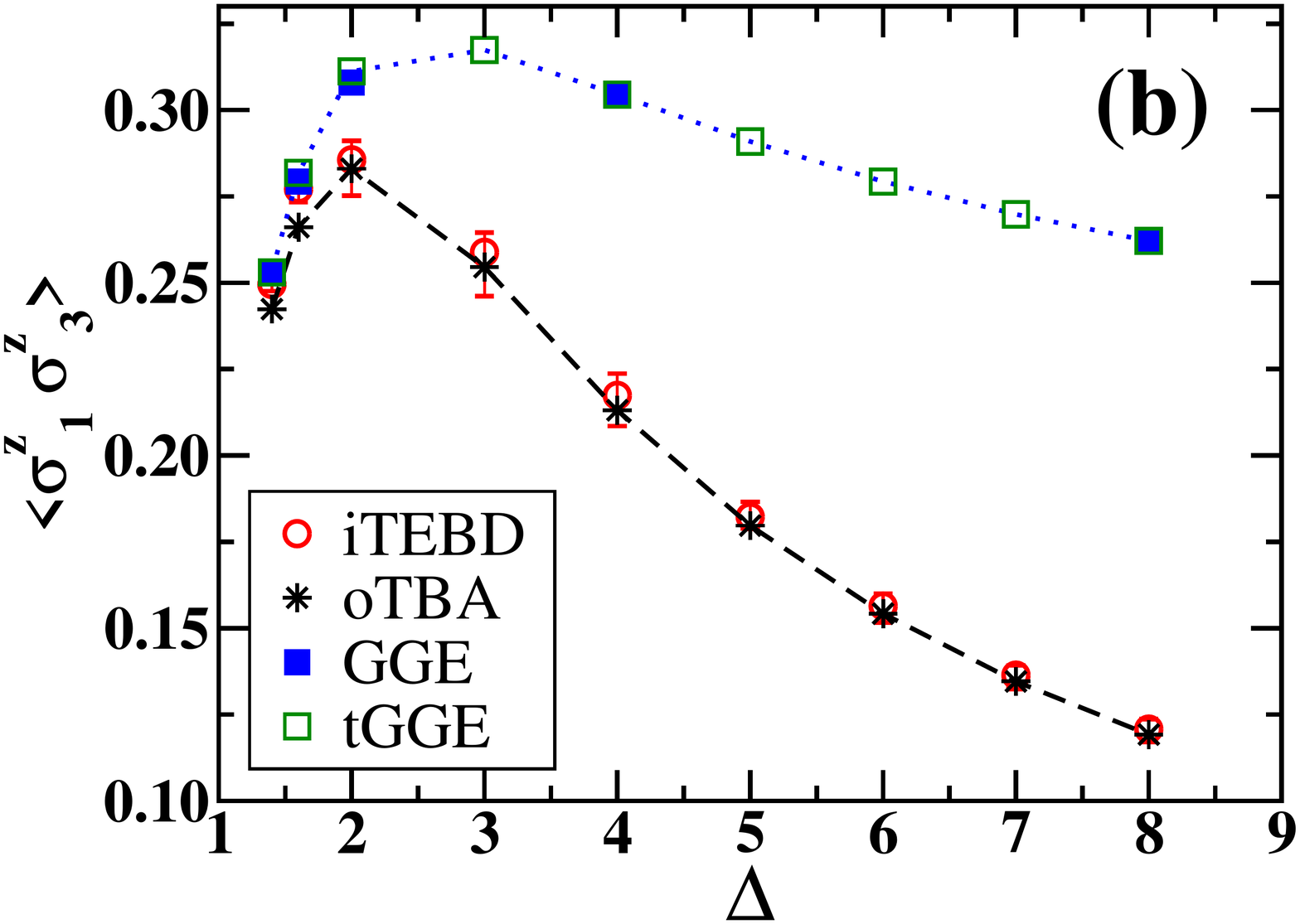}}
\caption{$\Delta$ dependence of the ITEBD (red circles with error bars), overlap TBA (black stars), full GGE \cite{2014PhRvB..89l5101F} (full blue squares) and truncated GGE (empty green squares) results for the large time expectation values (a) $\langle\sigma_1^z\sigma^z_2\rangle$ and (b) $\langle\sigma_1^z\sigma^z_3\rangle$ after the quench from the dimer initial state \erf{dimer}.} 
\label{fig2}
\end{figure*}


\paragraph{Failure of the GGE description of the steady state correlations.---}

To demonstrate that the GGE cannot give an appropriate description of the steady state, 
we compare  
 its predictions for correlation functions with the results of real-time 
 out-of-equilibrium numerical simulations, performed using 
 the infinite size time evolving block decimation (ITEBD) algorithm \cite{2004PhRvL..93d0502V,Vidal}.   
 In the simulations,  translational invariance of the initial states in Eqs. \erf{neel} and 
\erf{dimer} was implemented by averaging over the two components of the 
states~\cite{FootnoteTransInv}.
The time evolution of three different correlators is shown in Fig.\ \ref{fig1} for a quench starting from the dimer state  
for $\Delta=4$ (red lines). The correlation functions quickly converge to stationary values before 
 the simulations  break down for large times  (shaded regions)~\cite{FootnoteITEBDValid}.

Strikingly, the exact GGE values of Ref. \cite{2014PhRvB..89l5101F} (blue dotted lines in Fig.\ \ref{fig1})  deviate significantly 
from the ITEBD results.  We report similar deviations for all $\Delta>1$ in Fig. \ref{fig2}, where we display the long-time asymptotic values extracted from the ITEBD simulation (red circles) together with the GGE predictions (squares) for $\langle\sigma_1^z\sigma^z_2\rangle$ and $\langle\sigma_1^z\sigma^z_3\rangle$,  
as functions of $\Delta$.  We included truncated GGE (TGGE) results \cite{2013JSMTE..07..003P,2013JSMTE..07..012F} as well, obtained by keeping the first six nonzero charges in the density matrix \erf{eq:GGE}. 
The discrepancy between the GGE and the ITEBD results is evident (for additional numerical data, see the Supplemental Material \cite{SM}).

The mismatch between  real-time simulations and  GGE results could be, in principle, the result of long relaxation times beyond the reach of the ITEBD simulations. We rule out this possibility by applying an alternative theoretical method, the overlap thermodynamic Bethe ansatz (OTBA) which implements the quench-action approach  \cite{2014PhRvA..89c3601D,2013PhRvL.110y7203C} (see below), and predicts {\it exactly} the asymptotic values of the correlations. These values, shown as black dashed lines in Fig.\ \ref{fig1} and black stars in Fig.\ \ref{fig2}, are clearly in excellent agreement with our ITEBD results. 
The state predicted by the OTBA is a steady state of maximal conditional entropy.   
Therefore, the excellent agreement proves that the ITEBD simulation has reached the true asymptotic steady state, and, thus makes the evidence for the breakdown of the GGE conclusive. Moreover, it also demonstrates that the quench-action-approach-based OTBA indeed correctly captures the steady state of the \textit{XXZ} model.

The TGGE predictions evaluated in Ref. \cite{2013JSMTE..07..003P} and the OTBA results are also different
in the N\'eel case, but for large $\Delta,$ where the TGGE is reliable, this difference is beyond our numerical resolution (Supplemental Material \cite{SM}). 
We obtained similar results for $x$-$x$ correlators for both initial states, although for the latter ITEBD is not as accurate as for the $z$-$z$ 
correlators (cf. Ref. \cite{future}). 

We remark that translational invariance would be broken for the dimer state without the symmetrization under $\hat T$ in Eqs. \erf{neel} and \erf{dimer},   and it is yet an open question whether it is restored in the long-time limit after the quench \cite{2014PhRvB..89l5101F,2014JSMTE..03..016F}. 
Nevertheless, the GGE was expected to describe the average over the two components of the states \erf{neel} and \erf{dimer},
and yet it fails to do so. Note that for the observable $\langle\sigma_1^z\sigma^z_3\rangle$ translational averaging is immaterial 
since it is identical on both sublattices.

\paragraph{Overlap thermodynamic Bethe ansatz (OTBA).---} The OTBA method is formulated in the framework of the Bethe ansatz (BA) \cite{1931ZPhy...71..205B}, used to 
diagonalize the \textit{XXZ} Hamiltonian \erf{H}. Eigenfunctions corresponding to spin waves of $M$ flipped spins 
on a chain of length $L$ are parametrized by a set of $M$ complex numbers $\{\lambda_j\}$, called rapidities.
Spin waves can form bound states described by specific configurations of rapidities, called strings \cite{Takahashi2005Thermodynamics}.
In the thermodynamic limit (TDL), defined by $M\to\infty$, $L\to\infty$ with $M/L$ fixed,
it is convenient to introduce the densities $\rho_{n}(\lambda)$ of $n$ strings in rapidity space, together with the densities $\rho_{n}^{h}(\lambda)$ of unoccupied levels (holes). 

The quench-action approach of  Ref. \cite{2013PhRvL.110y7203C} has been developed
 to describe the  steady state following a quantum quench in Bethe ansatz integrable systems
and has been   implemented before for the transverse field Ising chain \cite{2013PhRvL.110y7203C} and the Lieb-Liniger model \cite{2014PhRvA..89c3601D}.
This variational approach  selects the relevant states by minimizing the generalized free energy 
(quench action)
\be
\mc{S}(\{\rho_{n}\})=-\frac2L\mathrm{Re}\ln\langle\Psi_0|\{\rho_{n}(\lambda)\}\rangle-s(\{\rho_{n}(\lambda)\}).
\label{Sq}
\ee
The first term involves the overlap between the initial state $|\Psi_0\rangle$ and the steady state $|\{\rho_{n}(\lambda)\}\rangle$, characterized by the string densities $\{\rho_n(\lambda)\}$. 
The exact overlaps were computed for the N\'eel state 
 \erf{neel}  in Refs.~\cite{karolos,2013arXiv1309.4593P,BrockmannGeneralizedGaudin2014B,BrockmannGeneralizedGaudin2014A}. 
Here we  also generalized the overlap formula for the dimer product state \erf{dimer} along the lines of Ref. \cite{2013arXiv1309.4593P}.
In the TDL, the logarithm of both overlaps can be written as 
\be
-\frac2L\mathrm{Re}\, \ln\langle\Psi_0|\{\rho_{n}(\lambda)\}\rangle=\sum_{n=1}^{\infty}\int_{-\pi/2}^{\pi/2}d\lambda\rho_{n}(\lambda)g_{n}(\lambda)
\label{overlap}
\ee
with the $g_{n}(\lambda)$ functions given by
\begin{subequations}
\label{gn}
\begin{align}
g_{1}^{\textit{N}}(\lambda)= & -\ln\frac{\tan(\lambda+\frac{i\eta}{2})\tan(\lambda-\frac{i\eta}{2})}{4\sin^{2}(2\lambda)},\label{gNeel}\\
g_{1}^{\textit{D}}(\lambda) =& -\ln\frac{\sinh^4(\eta/2)\cot^2(\lambda)}{\sin(2\lambda+i\eta)\sin(2\lambda-i\eta)}
\end{align}
\end{subequations}
for the N\'eel and dimer states, respectively. In both cases $g_{n}^{\textit{N,D}}(\lambda)=  \sum_{j=1}^{n}g_{1}^\textit{N,D}\left[\lambda+(i\eta/2)(n+1-2j)\right]$ for higher strings ($n>1$).

The second term $s(\{\rho_{n}(\lambda)\})$ in Eq.~\eqref{Sq} is the entropy density \cite{1969JMP....10.1115Y,Takahashi2005Thermodynamics} accounting for the number of microstates realizing the set of macroscopic $\{\rho_n(\lambda)\}$ 
\be
s(\rho_n)=\sum_{n=1}^{\infty}\int
d\lambda\,\frac{1}{2}\left[\rho_{n}\ln\frac{\rho_{n}+\rho_{n}^{\textrm{h}}}{\rho_{n}}+\rho_{n}^{\textrm{h}}\ln\frac{\rho_{n}+\rho_{n}^{\textrm{h}}}{\rho_{n}^{\textrm{h}}}\right].
\label{entropy}
\ee
It is exactly half of the standard Yang-Yang entropy density  \cite{1969JMP....10.1115Y,Takahashi2005Thermodynamics} due to the fact that only parity invariant microstates have nonzero overlap \cite{2014PhRvA..89c3601D,BrockmannGeneralizedGaudin2014A}. 

The quench action \erf{Sq} expresses the idea that the states relevant in the TDL are those with both large overlaps with the initial state and a large number of microscopic realizations. 
The steady state is captured by a saddle-point set of string densities,  with the saddle-point approximation becoming exact in the TDL. 
The densities $\{\rho_n(\lambda)\}$ and $\{\rho_{n}^{\textrm{h}}(\lambda)\}$ are, however, not independent;
interactions couple the rapidities of all the spin excitations,  as expressed by the Bethe equations. 
 As a consequence, the densities  $\{\rho_n\}$ and $\{\rho_{n}^{\textrm{h}}\}$  satisfy coupled integral equations (constraints) \cite{Takahashi2005Thermodynamics}
\be
a_{n}(\lambda)  =\rho_{n}(\lambda)  +\rho_{n}^{\text{h}}(\lambda)  +\sum_{m=1}^{\infty} [T_{nm}\!\circ\!\rho_{m}](\lambda).
\label{eq:BetheEquations}
\ee
Here, $[a\!\circ\!b](\lambda)= \int_{-\pi/2}^{\pi/2}d\lambda'a(\lambda-\lambda')b(\lambda')$ denotes convolution,  
and the interaction kernel is expressed as 
$
T_{nm} =(1-\delta_{n,m})a_{|n-m|}+a_{n+m}+2\sum_{j=1}^{\text{min}\{n,m\}-1} a_{|n-m|+2j}
$, with $\pi a_{n}(\lambda)=\sinh(n\eta)/[\cosh(n\eta)-\cos(2\lambda)]$ and  $\cosh\eta =\Delta$.
  The constrained extremum of $\mc{S}(\{\rho_{n}(\lambda)\})$ 
is then found through the standard treatment \cite{Takahashi2005Thermodynamics} and leads to the following
integral equations for the functions $\eta_{n}(\lambda)\equiv \rho_{n}^{\textrm{h}}(\lambda)/\rho_{n}(\lambda)$:
\be
\ln\eta_{n}(\lambda)= g_{n}(\lambda)+\mu n+\sum_{m=1}^{\infty}[T_{nm}\!\circ\ln(1+\eta_{m}^{-1})](\lambda).
\label{eq:GTBA}
\ee
Here,  $\mu$ is a Lagrange multiplier introduced to fix the overall magnetization to zero.
By examining the $g_n(\lambda),$ we obtained for both initial states  that $\ln\eta_n\sim \eta\,n^2$ for large $n.$
This implies that the higher strings are suppressed as $\propto e^{-\eta n^2}$, so the infinite set of equations \eqref{eq:GTBA} 
can be safely truncated to relatively few equations, and solved numerically.
Having the numerical solution for $\eta_n$ at hand, we can replace them into  Eq. \eqref{eq:BetheEquations} and determine 
the densities efficiently. 
(For further technical details see Ref. \cite{future}).

\paragraph{OTBA consistency checks. ---}
There are several ways to check that the numerically obtained saddle-point string densities $\{\rho^\ast_n\}$   
are, indeed, correct and correspond to the right initial states and to the right saddle point.
A nontrivial check is provided by the computation of the norm of the initial state in the TDL, i.e.,
\be
0 = -\frac1L \ln\langle\Psi_0|\Psi_0\rangle = \mc{S}(\{\rho^\ast_n(\lambda)\}),
\label{sumrule}
\ee
with the quench action Eq. \erf{Sq} evaluated via Eqs. (\ref{entropy}) and (\ref{overlap}) for the saddle-point solution. 
Equation~\eqref{sumrule} is, indeed, satisfied by our saddle-point solution within the accuracy of our numerical 
simulations $\mc{O}(10^{-8}).$ 
Notice that if this integral sum rule was violated, then the spectral weight of the saddle-point solution would be zero in the TDL.

Another important consistency check is provided by the expectation values of the conserved charges. 
These can be expressed in terms of the saddle-point densities as  
\cite{1976NuPhB.117..475L} 
\be
\langle \hat Q_{2m}\rangle = \sum_{n=1}^\infty \int_{-\pi/2}^{\pi/2}d\lambda \;\rho^\ast_n(\lambda)q^{(2m)}_n(\lambda),
\label{Qrho}
\ee
with $q^{(2m)}_{n}(\lambda) = 2\pi \sum_{j=1}^{n}[-(\p/\p\lambda)]^{2m} a_1[\lambda+(i\eta/2)(n + 1-2j)]$, 
the energy, in particular, being  given by $E=2\sinh\eta \langle \hat Q_{2}\rangle.$ The saddle-point values of these  
charges must equal their values in the initial states.  These latter were computed for the symmetric N\'eel state
 in Refs. \cite{2013JSMTE..07..003P,2013JSMTE..07..012F}. Here we determined them 
using both these methods  in the symmetrized dimer state, $|\Psi_0^D\rangle$.  We evaluated the first six nonzero charges, 
 $\{\langle\hat Q_{2m}\rangle\}_{m=1,\dots,6}$ and compared them with the expectation values computed from Eq. \erf{Qrho}.
Excellent  agreement  up to more than 8 digits is found in all cases, providing a further stringent verification of the OTBA solution.

\paragraph{Steady state correlations.---}
With the saddle-point string densities at hand, we then
computed various short distance correlation functions in the steady state by making use of the recent results of two of the present authors, who provided exact formulas for the 2-point correlation functions in terms of the string densities \cite{2014arXiv1405.0232M}. We compared these values with
the results of our  ITEBD simulations. Excellent agreement is found between  OTBA and ITEBD  for both initial states 
and for all $\Delta>1$ values (cf. Figs.\ \ref{fig1} and \ref{fig2} and the Supplemental Material \cite{SM} for detailed numerical data). 
This establishes the quench-action-approach-based OTBA as a correct description of the steady state
and the failure of GGE at the same time. 

\paragraph{Discussion.---}
In this Letter, we studied various correlation functions in the asymptotic steady states for quantum quenches in the \textit{XXZ} spin chain. 
We found that the  predictions of the generalized Gibbs ensemble differ significantly from the results of 
real-time ITEBD simulations in the dimer case, thereby signaling the breakdown of the GGE. 
We also determined these asymptotic correlators by applying the quench-action-based overlap TBA description of the steady state, 
and obtained numerically accurate predictions. We found that while  the quench-action-based OTBA correctly 
captures the asymptotic steady state, the GGE fails for the states considered here.
Finding a macroscopic statistical ensemble description of the steady state and clarifying 
the conditions for the validity of the GGE in strongly interacting systems, therefore, remain intriguing 
open questions.

\paragraph{Acknowledgments.---} 
We would like to thank F. Pollmann and P. Moca for numerous discussions, their valuable feedback and the help they gave us to set up the TEBD calculations. M.K. is grateful to J.-S. Caux, J. De Nardis, and B. Wouters for useful discussions on some aspects of their work. M.K. acknowledges financial support from the Marie Curie IIF Grant No. PIIF-GA-2012- 330076. M.A.W. and G.Z. acknowledge financial support from Hungarian Grants No. K105149 and CNK80991.

\paragraph{Note added.---} During the final stages of this work, a related paper appeared \cite{2014arXiv1405.0172W} that independently arrived at Eq. \erf{gNeel} and in which the difference between the GGE and the OTBA predictions for nearest-neighbor correlators in the quench starting from the N\'eel state was observed as well.

\pagebreak














\setcounter{equation}{0}%
\setcounter{figure}{0}%
\setcounter{table}{0}%
\renewcommand{\thetable}{S\arabic{table}}
\renewcommand{\theequation}{S\arabic{equation}}
\renewcommand{\thefigure}{S\arabic{figure}}

\onecolumngrid

\begin{center}
{\Large Supplementary Material for EPAPS \vspace{2mm}\\ 
Correlations after a quantum quench in the XXZ spin chain: failure
of the generalized Gibbs ensemble}
\end{center}

\section*{Numerical methods}

In this section we present a collection of high precision numerical data as well as some details on the numerical simulations.

\bigskip

\paragraph{ITEBD simulation.---}

We performed the simulations by an ITEBD-code, which made use of the $U(1)$ rotational symmetry of the XXZ model around the $z$-axis. Maximal bond dimensions of the $U(1)$ blocks were set to $\chi_{\mathrm{block}} = 400$, resulting in a total bond dimension $\chi_{\mathrm{tot}} > 1000$. We used a first order Trotter-expansion  for the time evolution operator with a time step $dt = 0.001$, and  verified that decreasing this time step does not   modify our results.

To control the reliability of our data, we computed the truncated weight at each time-step, and requested that
 the one-step truncated weight be less than $10^{-8}$. As an
 alternative method, we ran the code with different bond dimensions and determined  the time,
where the results deviated upon increasing bond dimension. The two methods resulted
in  approximately the same threshold time.  The error bars in Fig.\ 2 of the main text and Tables \ref{neeltab} and \ref{dimtab} below were then estimated by extracting the minimal  and maximal  values of the correlators  before the threshold time in the last time interval of length $1.0$.

For $\Delta=1.4$ and $\Delta=1.6$ it is difficult to estimate the error because for the times the simulation can reach there is still a drift of the correlation functions; in these cases the errors shown in
the tables are computed from the fluctuations around this drifting average. The remaining drift
of the average introduces a further systematical error, which is not included in our statistical estimate, and is hard to quantify reliably.

\bigskip

\paragraph{Precision of the OTBA solution.---}

The precision of the solution of the OTBA system was improved by extrapolation in the discretization used in the numerical solutions of the integral equations.

Checking the sum rule (11) by plugging in our numerical solution for different values of $\Delta$ for both initial states we found that while both the overlap (6) and the entropy (8) are of order $\mc{O}(10^{-1}),$ their difference, i.e. the quench action (5) is of order $\mc{O}(10^{-8})$ and shrinks while increasing the number of equations included,  therefore it is interpreted as a numerical error of the truncated OTBA system. This provides  very strong evidence that we indeed found the correct saddle point solution. 

\bigskip

\paragraph{Results.---}

The estimates of the asymptotic values of the correlations together with the OTBA predictions  are listed for various $\Delta$ values in Table~\ref{neeltab} for the zero momentum N\'eel state and in Table~\ref{dimtab}  for the dimer initial states. Based on the accuracy of the saturation of the sum rule and of the mean values of conserved charges, we can trust the results of the OTBA at least up to 7 digits. The tables also show the TGGE predictions and, when available from \cite{2014PhRvB..89l5101F}, the predictions of the full GGE as well.

In the N\'eel case the difference between the OTBA and TGGE results \cite{2013JSMTE..07..003P} is smaller than the accuracy of the ITEBD, except for $\Delta=2$ where the TGGE seems to fail. However, in this case the TGGE with 6 charges has not yet converged in the truncation level. We expect that in the N\'eel case only the quenches to $\Delta<1$ can conclusively decide between the OTBA and GGE predictions.

\begin{table}[h!]
  \centering
  \begin{tabular}{|c||c|c|c|}
    \hline
 & $\vev{\sigma_1^z\sigma_2^z}$ & $\vev{\sigma_1^z\sigma_3^z}$ &
$\vev{\sigma_1^z\sigma_4^z}$   \\ 
\hline\hline
$\Delta=2$, ITEBD&-0.6623(37) &0.3963(68) &-0.2967(50)  \\
\hline
 OTBA &  -0.6602541 & 0.3929011& -0.2921120  \\
\hline
TGGE & -0.648906 & 0.386578 & -0.280383 \\
\hline\hline
$\Delta=3$, ITEBD& -0.8145(28)	 &0.6493(52) &-0.5763(45)  \\
\hline
 OTBA &     -0.8151215 & 0.6480874 & -0.5745134  \\
\hline
TGGE &  -0.8140907 & 0.6501061 & -0.5757850 \\
\hline\hline
$\Delta=4$, ITEBD&-0.8875(18)	 &0.7815(34)&-0.7321(33)	 \\
\hline
 OTBA &  -0.8875741 & 0.7812765 & -0.7317703  \\
\hline
TGGE &  -0.8874100 & 0.7824718 & -0.7328197 \\
\hline\hline
$\Delta=5$, ITEBD & -0.9254(9)  & 	0.8537(16)  & -0.8193(16)\\
\hline
 OTBA &  -0.9253052 & 0.8532200 & -0.8187737  \\
\hline
TGGE &  -0.9252648 & 0.8538195 & -0.8193336 \\
\hline
  \end{tabular}
\caption{Correlation functions in the steady state computed from the real time ITEBD simulation starting from the N\'eel state, compared with the OTBA and TGGE predictions.}
\label{neeltab}
\end{table}

\begin{table}
  \centering
  \begin{tabular}{|c||c|c|c|}
    \hline
 & $\vev{\sigma_1^z\sigma_2^z}$ & $\vev{\sigma_1^z\sigma_3^z}$ &
$\vev{\sigma_1^z\sigma_4^z}$   \\ 
\hline
\hline
$\Delta=1.4$, ITEBD&-0.5668(8)&0.2495(16)&-0.1468(8) \\
\hline
OTBA & -0.5654103 &   0.2423652 & -0.1399874 \\
\hline
GGE & -0.5583 & 0.2531 & -0.1428 \\
\hline
TGGE &-05582489 &   0.2530204 &  -0.1426594  \\
\hline\hline
$\Delta=1.6$,  ITEBD & -0.5881(14)& 0.2771(35) &-0.1640(27)	 \\
\hline
OTBA &  -0.5844820 &  0.2660954&  -0.1523768 \\
\hline
GGE & -0.5751 & 0.2793 & -0.1575 \\
\hline
TGGE &   -0.5771005 &  0.2818048 & -0.1607061\\
\hline
\hline $\Delta=2$, ITEBD &-0.6017(36)&0.2856(79)&-0.1493(61) \\
\hline
 OTBA &
 -0.6028190&  0.2830639 & -0.1483621  \\
\hline
GGE & -0.5919  & 0.3080 & -0.1653 \\
\hline
TGGE &  -0.5943014 &  0.3112498&  -0.1694210  \\
\hline
\hline
$\Delta=3$, ITEBD &-0.6030(46)&0.2588(92)&-0.0922(75)\\
\hline
 OTBA &
-0.6028720 & 0.2545210& -0.0896056\\
\hline
TGGE &  -0.5945574 &  0.3173677 & -0.1409851 \\
\hline
\hline
$\Delta=4$, ITEBD &-0.5906(38)&0.2174(76)&-0.0421(56) \\
\hline
 OTBA &
-0.5900476 & 0.2130988& -0.0391867\\
\hline
GGE & -0.5842 & 0.3045 & -0.1118 \\
\hline
TGGE &  -0.5841759 &  0.3045031 & -0.1117982  \\
\hline
\hline
$\Delta=5$, ITEBD &-0.5782(21) &0.1823(38)&-0.0062(30)\\
\hline
 OTBA &
 -0.5780574& 0.1796867& -0.0048808  \\
\hline
TGGE &
 -0.5739028 &  0.2908932&  -0.0900247  \\
\hline
\hline
$\Delta=6$, ITEBD &-0.5684(21) &0.1564(41) & 0.0175(34) \\
\hline
OTBA &   -0.5683073 &   0.1541999 &  0.0187079\\
\hline
TGGE & -0.5652571 &   0.2793170 &  -0.0740645  \\
\hline
\hline
$\Delta=7$, ITEBD & -0.5606(18) & 0.1364(34) &0.0345(29) \\
\hline
OTBA &   -0.5605040  & 0.1345920  & 0.0355613  \\
\hline
TGGE &  -0.5581836 &   0.2698481 &  -0.0621113  \\
\hline
\hline
$\Delta=8$, ITEBD &-0.5543(16)	& 0.1208(33) &0.0473(26)  \\
\hline
OTBA &  -0.5542031 &  0.1191975 & 0.0480651  \\
\hline
GGE & -0.5524 & 0.2621 & -0.05291 \\
\hline
TGGE & -0.5523842 &  0.2621091 & -0.0529131  \\
\hline
  \end{tabular}
\caption{Correlation functions in the steady state computed from the real time ITEBD simulation starting from the dimer state, compared with the OTBA, TGGE and GGE predictions.}
\label{dimtab}
\end{table}



\begin{thebibliography}{10}
\bibitem{FootnoteTransInv}
This approach is justified in the TDL,   where off-diagonal contributions vanish.
\end{thebibliography}
\end{document}